\documentclass[
    ,final            
  ,numberedheadings 
  ]
  {aipproc}

\layoutstyle{6x9}

\begin{document}

\title{Hadronic decays of the tau lepton into $KK\pi$ modes within Resonance Chiral Theory\footnote{Poster presented at the International Workshop on Quantum Chromodynamics: Theory and Experiment (QCD@Work 2007), Martina Franca (Bari, Italy), 16th-20th June 2007; Reports: IFIC/07-56,FTUV/07-0924. To appear in the Proceedings.}}


\author{P.~Roig}{
  address={Instituto de F\'isica Corpuscular, IFIC, CSIC-Universitat de Val\`encia.\\ Apt. de Correus 22085, E-46071 Val\`encia, Spain}
}



\setlength{\oddsidemargin}{0.cm}
\setlength{\evensidemargin}{0.cm}
\setlength{\textwidth}{15cm}
\setlength{\topmargin}{-0.5cm}
\setlength{\textheight}{22.5cm}
\setlength{\headsep}{1cm}

\def\onebox{{\vbox{\hbox{$\sqr\thinspace$}}}}
\def\twobox{{\vbox{\hbox{$\sqr\sqr\thinspace$}}}}
\def\threebox{{\vbox{\hbox{$\sqr\sqr\sqr\thinspace$}}}}
\def\N{N_C}
\def\Tr{{\rm Tr}}
\def\CPT{\chi PT}
\def\RCPT{R\chi T}
\def\RCT{R\chi T}
\def\t{\tau}
\def\mapright#1#2{\smash{
     \mathop{-\!\!\!-\!\!\!\rightarrow}\limits^{#1}_{#2}}}
\def\p{\pi}
\def\m{\mu}
\def\n{\nu}
\def\vr{\varrho}
\def\r{\rho}
\def\s{\sigma}
\def\e{\eta}
\def\g{\gamma}
\def\l{\lambda}
\def\vf{\varphi}
\def\f{\phi}
\def\vf{\varphi}
\def\ep{\epsilon}
\def\vep{\varepsilon}
\def\b{\beta}
\def\a{\alpha}
\def\d{\delta}

\providecommand{\openone}{\leavevmode\hbox{\small1\kern-3.8pt\normalsize1}}

\newcommand{\cO}{{\cal O}}
\newcommand{\ket}{\,\rangle}
\newcommand{\bra}{\langle \,}

\begin{abstract}
\noindent$\tau$ decays into hadrons have a twofold interest: On the one hand, they are a clean environment for studying the hadronization of the left-handed current of QCD, while, on the other side, provide relevant dynamical information of the resonances that mediate these processes. Within an effective field theory-like framework, namely Resonance Chiral Theory, we analyse the decays ot the $\t$ into $KK\pi$ modes and compare the results with CLEO and BaBar data. In this way, we provide bounds on the couplings entering our Lagrangian and predict the corresponding spectral functions. As a main result -and contrary to the bulk of theoretical studies and experimental analyses- we find vector current dominance on these decays.\\
%
\end{abstract}

\maketitle


\section{INTRODUCTION} \label{intro}
Although it is perfectly known that Quantum Chromodynamics \cite{QCD} (QCD) is the quantum field theory that describes the strong interaction, we do not know how to solve it when it becomes non-perturbative. That happens, roughly speaking, below 2 GeV. Ideally, one would like to complement partonic QCD with a dual theory written in terms of the relevant degrees of freedom in the low and intermediate energy domain (mesons and baryons) that could be treated perturbatively at large distances, easening the computations and ensuring convergence. Unfortunately, it is not yet known how to accomplish this task.\\
\indent $\t$ decays into hadrons allow to study the hadronic properties of vector and axial-vector QCD currents and, accordingly, to find the dynamics generated by the exchanged resonances. At very low energies ($E\ll M_\rho$), Chiral Perturbation Theory ($\chi$PT) \cite{Weinberg:1978kz, CPT} is the effective field theory of QCD, but only enables to explain hadronic tau decays in a tiny window of the available phase space \cite{Colangelo:1996hs}. However, we cannot neglect the importance of the resonance states exchanged in the process. Applying vector meson dominance \cite{VMD}, the most relevant resonances are the $\rho$(770), the $\omega$(782), the $K^*$(892) and the $a_1$(1260). To explore them, one should include these degrees of freedom explicitly into the theory.\\
\indent There have been several strategies to tackle the problem: the best-known one is that of employing a parameterization using a combination of Breit-Wigner's \cite{VMD, Kuhn:1990ad}. Even though it is clear that polology demands 'something of the kind', it is not clear how to go beyond and formalize its link with QCD, if it exists. In fact, the analysis of the decay $\t\to3\pi\n_\t$ undertaken in Refs.
~\cite{Portoles:2000sr, GomezDumm:2003ku} showed that this procedure violates the chiral symmetry of massless QCD at $\cO(p^4)$.\\
\indent Several experiments have been collecting good quality data on $\t\to KK \pi\n_\t$ decays. In particular, CLEO and BaBar have published the rates for these channels.\\
\indent It is known \cite{Liu:2002mn} that the decay $\t^- \to K^+ K^- \pi^- \n_\t$ is not well described by a model generalizing the one given in Ref.~\cite{Kuhn:1990ad}. As a result of this, the CLEO collaboration proposed \cite{Coan:2004ep} a modification of it, able to fit the data at the price of violating the Wess-Zumino normalization directly stemming from the chiral anomaly of QCD, as was put forward in Ref. \cite{Portoles:2004vr}.\\
\indent All these facts reinforce the desirability of a less model-dependent analysis of hadronic $\t$ decays in order to gain more insight on the hadronization of the relevant QCD currents.\\
\section{THE RESONANCE CHIRAL THEORY OF QCD} \label{RChT}
As the energy increases approaching $M_\r$, it becomes unavoidable to explicitly introduce the new active degrees of freedom (the resonance mesons) into the effective action. The procedure, developed in Refs. \cite{Ecker:1988te, Ecker:1989yg} and known as Resonance Chiral Theory ($\RCT$), is ruled by the approximate chiral symmetry of low-energy QCD for the lightest pseudoscalar mesons and by unitary symmetry for the resonances. Its two theoretical pillars are:
\begin{itemize}
\item The Weinberg's Theorem \cite{Weinberg:1978kz} that ensures that writing the most general Lagrangian consistent with the assumed symmetries we will obtain the most general $S$-matrix elements respecting analiticity, perturbative unitarity, cluster decomposition and the starting symmetry principles and, in particular, local chiral symmetry in the presence of external fields \cite{Leutwyler:1993iq}.
\item The large-$N_C$ expansion of QCD. It has been suggested \cite{Nc} that $1/N_C$ ($N_C$ is short for the number of colors) could serve as a perturbative expansion parameter. Many features of meson phenomenology find their explanation in the $N_C\to\infty$ framework \cite{Pich:2002xy} which supports the procedure. Relevant consequences of it are that meson dynamics is described, at LO in the expansion, by tree level diagrams given by an effective local Lagrangian including the interactions amidst an infinite number of zero-width resonances. However in most processes, like $\t$ decays, we need to include finite widths (a NLO effect in the $1/N_C$ expansion) as we do within our framework \cite{GomezDumm:2000fz}. We are also departing from the $N_C\to\infty$ limit because we consider just one multiplet of resonances per set of quantum numbers (single resonance approximation \cite{SRA}) and not the infinite tower predicted.
\end{itemize}
\indent Once we have highlighted the key points of our study, it is time to give some details about the concrete way we proceed. The relevant part of the $\RCT$ Lagrangian is \cite{Ecker:1988te, Cirigliano:2004ue, Ruiz-Femenia:2003hm, tesina}:
\begin{eqnarray} \label{Full_Lagrangian}
\mathcal{L}_{\RCT}=\frac{F^2}{4}\bra u_\m u^\m +\chi_+ \ket+\frac{F_V}{2\sqrt{2}}\bra V_{\mu\nu} f^{\mu\nu}_+ \ket +
\frac{i G_V}{\sqrt{2}} \bra V_{\mu\nu} u^\mu u^\nu\ket + \frac{F_A}{2\sqrt{2}}\bra A_{\mu\nu} f^{\mu\nu}_- \ket \nonumber\\
+\mathcal{L}_{\mathrm{kin}}^V+\mathcal{L}_{\mathrm{kin}}^A + \sum_{i=1}^{5}\lambda_i\mathcal{O}^i_{VAP} + \sum_{i=1}^7\frac{c_i}{M_V}\mathcal{O}_{VJP}^i +\sum_{i=1}^4d_i\mathcal{O}_{VVP}^i\sum_{i=1}^5 + \frac{g_i}{M_V} {\cal O}^i_{VPPP} \, ,
\end{eqnarray}
where all couplings are real, being $F$ the pion decay constant in the chiral limit. The notation is that of Ref.~\cite{Ecker:1988te}. Here and in the following $P$ stands for the lightest pseudoscalar mesons. Furthermore, all couplings in the second line are defined to be dimensionless. For the explicit form of the operators in the last line, see \cite{Ruiz-Femenia:2003hm, tesina, articulo}. In this work we have considered for the first time the $VPPP$ vertex in the odd-intrinsic parity sector.
 Notice that we are using the antisymmetric tensor formalism to describe the spin one resonances. We are not considering the $\cO(p^4)$ Lagrangian of Goldstone bosons to avoid double counting, because the LEC's of $\cO(p^4)$ are known to be saturated by resonance exchange \cite{Ecker:1989yg}.\\
\indent An all-important caveat, however, is that $\mathcal{L}_{\RCT}$ is not QCD for arbitrary values of its couplings. Hence, if we want to comprehend more about the non-perturbative behaviour of the strong interaction, we ought to learn about the determination of the couplings from the underlying theory \cite{Pich:2002xy}. Guided by this purpose we will demand several known features of QCD to our effective Lagrangian.\\
\indent  A matching procedure has been put forward \cite{GomezDumm:2003ku, Ecker:1989yg, Cirigliano:2004ue, Amoros:2001gf, Knecht:2001xc, Cirigliano:2006hb} by analysing (three-point) Green functions that are order parameters of massless QCD and matching the resonance theory with the LO in their OPE expansion. Additionally, if necessary, we will demand to the form factors we obtain a Brodsky-Lepage-like behaviour both to the vector and axial-vector ones.\\
\indent
 Phenomenology could also help us to fix bounds on the coupling in Eq. (\ref{Full_Lagrangian}). For instance, $F_V$ could be extracted from the measured $\Gamma(\r^0\to e^+e^-)$, $G_V$ from $\Gamma(\r^0\to \p^+\p^-)$, $F_A$ from $\Gamma(a_1\to \p\g)$ and the $\l_i$'s from $\Gamma(a_1\to \r\p)$ which starrs on the $\t\to3\p\n_\t$ processes themselves. $\Gamma(\omega\to \p\g)$, $\Gamma(\omega\to3\p)$ and the $\cO(p^6)$ correction to $\Gamma(\p \to \g\g)$ may give us information on the remaining couplings \cite{Ruiz-Femenia:2003hm}.\\
\section{FORM FACTORS IN $\tau^-\to(KK\pi)^-\,\nu_\tau$} \label{FormFactors}
The decay amplitudes for the processes $\t^-\to K^+\,K^-\,\pi^-\,\nu_\t$, $\t^-\to K^0\,\overline{K}^0\,\pi^-\,\nu_\t$ and $\t^-\to K^-\,K^0\,\pi^0\,\nu_\t$ can be written as
\begin{equation} \label{Mgraltau}
\mathcal{M}\,=\,-\frac{G_F}{\sqrt{2}}\,V_{us}\overline{u}_{\nu_\tau}\gamma^\mu(1-\gamma_5)u_\tau \mathcal{H}_\mu\,,
\end{equation}
where
\begin{eqnarray} \label{Hadronic_matrix_element}
\mathcal{H}_\mu & = & \bra P(p_1) P(p_2) P(p_3)|\left( \mathcal{V}_\m - \mathcal{A}_\m\right)  e^{i\mathcal{L}_{QCD}}|0\ket=\\
& & V_{1\mu} F_1^A(Q^2,s_1,s_2) + V_{2\mu} F_2^A(Q^2,s_1,s_2) + Q_\mu F_3^A(Q^2,s_1,s_2) + i V_{3\m} F_4^V(Q^2,s_1,s_2)\,,\nonumber
\end{eqnarray}
and
\begin{eqnarray}
V_{1\mu}\, = \, \left( g_{\mu\nu} - \frac{Q_{\mu}Q_{\nu}}{Q^2}\right) \,
(p_1 - p_3)^{\nu} \;\;\;\; , \;\;\;\; V_{2\mu}\, = \, \left( g_{\mu\nu} - \frac{Q_{\mu}Q_{\nu}}{Q^2}\right) \,
(p_2 - p_3)^{\nu}\,,\nonumber\\
V_{3\m} = \varepsilon_{\mu\nu\varrho\sigma}p_1^\nu\, p_2^\varrho\, p_3^{\sigma}\;\; \;\; , \;\;\;\; Q_\mu\,=\,(p_1\,+\,p_2\,+\,p_3)_\mu\;\;\;\;,\;\;\;\;s_i = (Q-p_i)^2\,.
\end{eqnarray}
Here $F_i$, $i=1,2,3$ correspond to the axial-vector current while $F_4$ drives the vector current. The form factors $F_1$ and $F_2$ have a transverse structure in the total hadron momenta, $Q^\m$, and drive a $J^P=1^+$ transition. The scalar form factor, $F_3$, vanishes with the mass of the Goldstone bosons (chiral limit) and, accordingly, gives a tiny contribution. In \cite{GomezDumm:2003ku} the vector form factor did not contribute in the isospin limit, so the aim of our work is not only to confirm or refuse the bounds on the $\lbrace \lambda_i\rbrace_{i=1}^5$ in that reference, but to explore the vector currect sector of the resonance Lagrangian.
\indent For details on the particular shape of the form factors, see Refs. \cite{tesina,articulo}.\\
\section{ASYMPTOTIC BEHAVIOUR AND QCD CONSTRAINTS} \label{Asymptotic_behaviour}
The explicit computation of the Feynman diagrams involved shows that the result depends only on three combinations of the $\lbrace \lambda_i\rbrace_{i=1}^5$, four of the $\lbrace c_i\rbrace_{i=1}^7$, two of the $\lbrace d_i\rbrace_{i=1}^4$ and three of the $\lbrace g_i\rbrace_{i=1}^5$. The number of free parameters has been reduced from 24 to 15.\\
\indent We require the form factors of the $A^\m$ and $V^\m$ currents into $KK\p$ modes vanish at infinite transfer of momentum. As a result, we obtain these constraints from $A^\m$:
\begin{eqnarray}
& & F_V\,G_V \, = \, F^2 \;\;\;\;\;\;,\;\;\;\;\;\;F_V^2-F_A^2 \, = \, F^2\;\;\;\;\;\;,\;\;\;\;\;\;M_V^2\,F_V^2 \, = \, M_A^2\,F_A^2\,,\nonumber\\
& & \lambda' \, = \, \frac{1}{2 \sqrt2}\; \frac{1}{\sqrt{1-\frac{F^2}{F_V^2}}} \sim 0.455\,\;\;\;\;,\;\;\;\;
\lambda'' \,  = \, - \left(1-
\frac{2\, F^2}{F_V^2}\right) \lambda'\sim-0.094\;, \label{constA}
\end{eqnarray}
contrary to $\lambda' = \frac{1}{2}$, $\lambda'' = 0$ and $F_V\,=\,\sqrt{2}\,F$ in \cite{GomezDumm:2003ku}. Furthermore, we have checked that the set of relations (\ref{constA}) are the most general ones satistying the demanded asymptotic behaviour in $\t\to3\p\n_\t$.\\
\indent Incidentally, the value of the ratio $\frac{F^2}{F_V^2}$ can be read from \cite{Cirigliano:2004ue} through:
$\frac{F_V^2}{F^2}\,=\,\frac{M_A^2}{M_A^2-M_V^2}\,$,
which yields -with $M_A\,=\,998(49)$MeV \cite{Mateu:2007tr}- $F_V\sim0.147$ GeV. Only $\l_0$ remains free. It was fitted to be around $12$ in Ref. \cite{GomezDumm:2003ku} and derived to be some $\frac{1}{8}$ in Ref. \cite{Cirigliano:2004ue}. A fit to  $\Gamma\left(\t\to K^+ K^- \p^-\n_\t\right)$ excludes the former value and favours the second one.\\
\indent Proceeding analogously for the vector current form factor, we find that it is only possible to satisfy the QCD requirements provided we include the $VPPP$ part of the Lagrangian in (\ref{Full_Lagrangian}). We have found the constraints \footnote{The folllowing definitions have been used: $c_{1235} \equiv c_1+c_2+8c_3-c_5$, $c_{125} \equiv c_1 - c_2 +c_5$, $c_{1256}\equiv c_1-c_2-c_5+2c_6$, $g_{123} \equiv g_1+2g_2-g_3$ and $d_{123} \equiv d_1 + 8d_2-d_3$.}
\begin{eqnarray}
\label{constraintsVFF}
c_{1256} & = & - \, \frac{N_C}{96 \pi^2} \, \frac{M_V F_V}{\sqrt{2} \, F^2}\sim -3.0\cdot10^{-2} (-3.5\cdot10^{-2})\; , \nonumber \\
d_3 & = & - \, \frac{N_C}{192 \pi^2} \frac{M_V^2}{F^2} \sim -0.11 (-0.08)\, , \nonumber  \\
g_2\, & = & \, \frac{N_C}{192 \pi^2}\, \frac{M_V}{\sqrt{2} \,F_V} \sim -5.9\cdot10^{-3}\; ,
\end{eqnarray}
as well as $c_{125} =  0 (0)$ and $g_{123} = 0$, where the values in parenthesis refer to the relations in Ref. \cite{Ruiz-Femenia:2003hm}. Although obtained in different kind of processes, they are quite similar. This points out to an error associated to the staying at $LO$ in the large-$N_C$ expansion (much) lower than the conservative $30\%$ usually claimed. From the 24 initially unknown couplings in Eq. (\ref{Full_Lagrangian}), only five remain free: $c_4$, $c_{1235}$, $d_{123}$, $g_4$ and $g_5$.\\
\indent We have also completed the computation of $\omega\to3\p$ in Ref. \cite{Ruiz-Femenia:2003hm} by including the local contribution. A fit to the branching ratio \cite{PDG2006} fixes $2g_4+g_5=-0.48\pm 0.01$, so only four parameters remain. In Ref. \cite{Ruiz-Femenia:2003hm} the relations $d_{123}=0.05$ and $c_{1235}=0$ were found. We have succeded in fitting the branching ratios using these values.\\
\indent These way, only $c_4$ and a combination of $g_4$ and $g_5$ remain to be known. In $\t^-\to K^-K^0\p^0\n_\t$, the computation of the Wess-Zumino term gives zero and the $VPPP$ one vanishes when imposing QCD short-distance behaviour. A fit to the branching ratio \cite{PDG2006} gives $c_4 = 0.06\pm0.01$. This lack of two contributions translates into a lower branching ratio for this channel. Keeping this value, we favour $g_4\sim-0.24$ and $0\sim g_5\ll g_4$. Remarkably, despite starting with 24 unknown couplings, we have been able to fix bounds on all entering the $\t\to KK\p\n_\t$ decays.\\
\section{RESULTS AND CONCLUSIONS} \label{Conclusions}
\begin{figure}
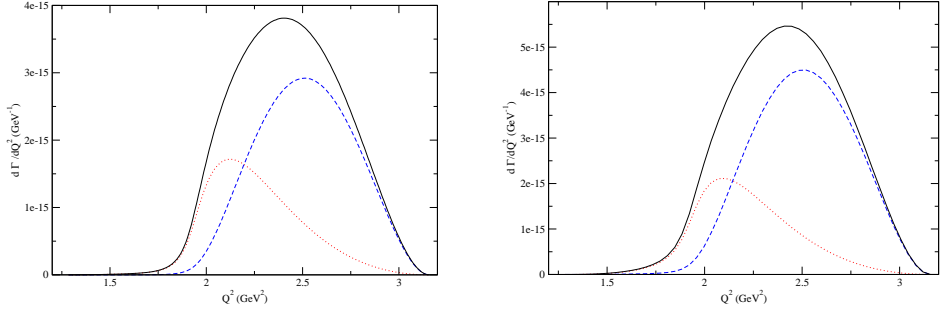
 \label{Fig}
\includegraphics[height=.18\textheight]{fig1.eps}\qquad
\includegraphics[height=.18\textheight]{fig2.eps}
\caption{Spectral function plotted versus the momentum of the hadronic system squared for the channels $
\tau^{-} \rightarrow K^{-}\, K^{0} \, \pi^{0}\, \nu_{\tau}$ (left), and $\tau^{-} \rightarrow K^{+}\, K^{-} \, \pi^{-}\, \nu_{\tau}$ (right). The dotted line corresponds to the axial-vector current contribution and the dashed line to the vector current one. The total contribution is plotted in solid line.}
\end{figure}

As a result of the procedure described above, we have obtained the branching ratios shown in Table \ref{Table}.\\
\begin{table}
\begin{tabular}{lrrrrrr}
\hline
Contribution/Source  & $\Gamma \left( \t^-\to K^+ K^- \p^- \n_\t\right)$
  &  $\Gamma\left( \t^-\to K^- K^0 \p^0 \n_\t\right)$  \\
\hline
\\Vector current, $\Gamma^V$ & $3.3\;(0.4)$ & $2.0\;(0.3)$
\\Axial-vector current, $\Gamma^A$ & $1.2\;(0.5)$ & $1.0\;(0.4)$
\\Total & $4.5\;(0.6)$ & $3.0\;(0.5)$\\
\hline
PDG'06 \cite{PDG2006} & $3.465\;(227)$ & $3.488\;(454)$\\
BaBar'07 \cite{:2007mh} & $3.048\;(084)$ &\\
CLEO'03 \cite{Liu:2002mn} & $3.511\;(136)$ &\\
\end{tabular}
\caption{\small{Comparison of our work with the branching ratios published by the experimental collaborations in units of $10^{-15}$ GeV. Our estimates for the errors are conservative educated guesses.}}
\label{Table}
\end{table}
\noindent Several comments are pertinent:
\begin{itemize}
\item Within the very accurate $SU(2)$ isospin limit, the following equality holds: $\Gamma\left( \t^-\to K^+ K^- \p^- \n_\t\right) = \Gamma\left( \t^-\to K^0 \overline{K}^0 \p^0 \n_\t\right)$, while PDG06 reports the value $\Gamma\left( \t^-\to K^0 \overline{K}^0 \p^0 \n_\t\right)\,=\,3.624\;(704)\cdot 10^{-15}$ GeV. Taking this and the values presented in Table \ref{Table} into account, it is probably safer to keep a central value close to $3.5\cdot10^{-15}$ GeV but enlarge the error to some $0.5\cdot10^{-15}$ GeV. The results from BaBar \cite{:2007mh} and CLEO \cite{Liu:2002mn} seem quite optimistic despite being rather different.\\
\item Our results are consistent with the PDG values \cite{PDG2006}. As already pointed out in \cite{Gomez-Cadenas:1990uj} we observe a clear dominance of the vector current, which is at odds with the conclusions given in Refs.~\cite{Coan:2004ep, Finkemeier:1995sr}.\\
\item As a new result, we find more decays to the kaon-antikaon channels. Schematically, we can summarize our findings as:
 \begin{equation}
 \frac{\Gamma^V}{\Gamma^A}\sim2 \;\;\;\;, \;\;\;\;\frac{\Gamma\left( \t^-\to K^+ K^- \p^- \n_\t\right)}{\Gamma\left( \t^-\to K^- K^0 \p^0 \n_\t\right)}\sim 1.5\,.
 \end{equation}
\item Our prediction for the spectral functions can be seen in Fig.~\ref{Fig}.\\
\end{itemize}
\indent This work is framed into a wider project that extends to all three meson decay channels of the $\t$. We are confident that new data from CLEO, BaBar, BELLE and a future super-B factory might allow us to revisit and improve the works done and to help the completion of all others.\\


\begin{theacknowledgments}
\noindent P.R. thanks Pietro Colangelo and Fulvia De Fazio for the remarkable organization and charming atmosphere of the QCD@Work 2007 meeting in Martina Franca (Italy). This work has been done in collaboration with D. G\'omez-Dumm, A. Pich and J. Portol\'es. I wish to thank the last one for a careful revision of the manuscript. P.R. is supported by a FPU contract (MEC). This work has been supported in part by the EU MRTN-CT-2006-035482 (FLAVIAnet), by MEC (Spain) under grant FPA2004-00996 and by Generalitat Valenciana under grants GVACOMP2007-156.\\
\end{theacknowledgments}


\begin{thebibliography}{9}

\bibitem{QCD}
H.~Fritzsch and M.~Gell-Mann,
  eConf {\bf C720906V2} (1972) 135
  [arXiv:hep-ph/0208010].
H.~Fritzsch, M.~Gell-Mann and H.~Leutwyler,
Phys.\ Lett.\ B {\bf 47} (1973) 365.


\bibitem{Weinberg:1978kz}
S.~Weinberg,
Physica A {\bf 96} (1979) 327.

\bibitem{CPT}
J.~Gasser and H.~Leutwyler,
Annals of Phys.\  {\bf 158} (1984) 142.
Nucl.\ Phys.\ B {\bf 250} (1985) 465.

\bibitem{Colangelo:1996hs}
G.~Colangelo, M.~Finkemeier and R.~Urech,
Phys.\ Rev.\ D {\bf 54} (1996) 4403
[arXiv:hep-ph/9604279].

\bibitem{VMD}
H.~K.~K\"uhn and F.~Wagner,
Nucl.\ Phys.\ B {\bf 236} (1984) 16.
A.~Pich,
In Proceedings of the {\it Tau Charm Factory Workshop} Ed.~L.~V.~Beers, SLAC (1989).

\bibitem{Kuhn:1990ad}
J.~H.~K\"uhn and A.~Santamar\'\i{}a,
Z.\ Phys.\ C {\bf 48} (1990) 445.

\bibitem{Portoles:2000sr}
  J.~Portol\'es,
  Nucl.\ Phys.\ Proc.\ Suppl.\  {\bf 98} (2001) 210
  [arXiv:hep-ph/0011303].

\bibitem{GomezDumm:2003ku}
D.~G\'omez Dumm, A.~Pich and J.~Portol\'es,
Phys.\ Rev.\ D {\bf 69} (2004) 073002
[arXiv:hep-ph/0312183].

\bibitem{Liu:2002mn}
F.~Liu  [CLEO Collaboration],
eConf {\bf C0209101} (2002) TU07
[Nucl.\ Phys.\ Proc.\ Suppl.\  {\bf 123} (2003) 66]
[arXiv:hep-ex/0209025].

\bibitem{Coan:2004ep}
T.~E.~Coan {\it et al.}  [CLEO Collaboration],
Phys.\ Rev.\ Lett.\  {\bf 92} (2004) 232001
[arXiv:hep-ex/0401005].

\bibitem{Portoles:2004vr}
J.~Portol\'es,
Nucl.\ Phys.\ Proc.\ Suppl.\  {\bf 144} (2005) 3
[arXiv:hep-ph/0411333].

\bibitem{Ecker:1988te}
G.~Ecker, J.~Gasser, A.~Pich and E.~de Rafael,
Nucl.\ Phys.\ B {\bf 321} (1989) 311.

\bibitem{Ecker:1989yg}
G.~Ecker, J.~Gasser, H.~Leutwyler, A.~Pich and E.~de Rafael,
Phys.\ Lett.\ B {\bf 223} (1989) 425.

\bibitem{Leutwyler:1993iq}
  H.~Leutwyler,
  Annals of Phys.\  {\bf 235} (1994) 165
  [arXiv:hep-ph/9311274].

\bibitem{Nc}
G.~'t Hooft,
Nucl.\ Phys.\ B {\bf 75} (1974) 461.
E.~Witten,
Nucl.\ Phys.\ B {\bf 160} (1979) 57.

\bibitem{Pich:2002xy}
  A.~Pich,
in Proceedings in the Phenomenology of Large $N_C$ QCD, edited by R.~Lebed (World Scientific, Singapore, 2002), p.239,
  [arXiv:hep-ph/0205030].

\bibitem{GomezDumm:2000fz}
D.~G\'omez Dumm, A.~Pich and J.~Portol\'es,
Phys.\ Rev.\ D {\bf 62} (2000) 054014
[arXiv:hep-ph/0003320].

\bibitem{SRA}
  M.~Knecht, S.~Peris, M.~Perrottet and E.~de Rafael,
  Phys.\ Rev.\ Lett.\  {\bf 83} (1999) 5230
  [arXiv:hep-ph/9908283].
  JHEP {\bf 0211} (2002) 003
  [arXiv:hep-ph/0205102].

\bibitem{Cirigliano:2004ue}
V.~Cirigliano, G.~Ecker, M.~Eidem\"uller, A.~Pich and J.~Portol\'es,
Phys.\ Lett.\ B {\bf 596} (2004) 96
[arXiv:hep-ph/0404004].

\bibitem{Ruiz-Femenia:2003hm}
P.~D.~Ruiz-Femen\'\i{}a, A.~Pich and J.~Portol\'es,
JHEP {\bf 0307} (2003) 003
[arXiv:hep-ph/0306157].

\bibitem{tesina}
P.~Roig,
Treball d'Investigaci\'o, Universitat de Val\`encia, 2006.

\bibitem{articulo}
P.~Roig, D.~G\'omez-Dumm, A.~Pich, J.~Portol\'es,
work in preparation, 2007.

\bibitem{Amoros:2001gf}
  G.~Amor\'os, S.~Noguera and J.~Portol\'es,
  Eur.\ Phys.\ J.\  C {\bf 27} (2003) 243
  [arXiv:hep-ph/0109169].

\bibitem{Knecht:2001xc}
  M.~Knecht and A.~Nyffeler,
  Eur.\ Phys.\ J.\  C {\bf 21} (2001) 659
  [arXiv:hep-ph/0106034].

\bibitem{Cirigliano:2006hb}
  V.~Cirigliano, G.~Ecker, M.~Eidem\"uller, R.~Kaiser, A.~Pich and J.~Portol\'es,
  Nucl.\ Phys.\  B {\bf 753} (2006) 139
  [arXiv:hep-ph/0603205].


\bibitem{Mateu:2007tr}
  V.~Mateu and J.~Portol\'es,
  Eur.\ Phys.\ J.\ C \textbf{52} (2007) 325-338
  [arXiv:hep-ph/0706.1039] .

\bibitem{PDG2006}
W.-M. Yao et al.,
Journal of Physics G \textbf{33}, 1 (2006)

\bibitem{:2007mh}
  B.~Aubert {\it et al.}  [BABAR Collaboration],
  arXiv:0707.2981 [hep-ex].

\bibitem{Gomez-Cadenas:1990uj}
J.~J.~G\'omez-Cadenas, M.~C.~Gonz\'alez-Garc\'\i{}a and A.~Pich,
Phys.\ Rev.\ D {\bf 42} (1990) 3093.

\bibitem{Finkemeier:1995sr}
M.~Finkemeier and E.~Mirkes,
Z.\ Phys.\ C {\bf 69} (1996) 243
[arXiv:hep-ph/9503474].

\end{thebibliography}
\end{document}